\documentclass[namedreferences]{solarphysics}
\usepackage[hyperref,optionalrh,solaromanenum]{spr-sola-addons} % For Solar Physics 
\usepackage{graphicx}                    % For eps figures, newer & more powerfull
\usepackage{color}                       % For color text: \color command
\usepackage{breakurl}                         % For breaking URLs easily trough lines
\begin{document}

\begin{article}

\begin{opening}

\title{Contribution to the Solar Mean Magnetic Field from Different Solar Regions}

%%%%%%%%%%%%%%%%%%%%%%%%%%%%%%%%%%%%%%%%%%%%%%%%%%%
%% Authors Names
%
\author[addressref=aff1,corref, email={alex.s.kutsenko@gmail.com}]{\inits{A.S.}\fnm{A.S.}~\lnm{Kutsenko}\orcid{0000-0002-1196-5049}}%\sep
\author[addressref={aff1,aff2}]{\inits{V.I.}\fnm{V.I.}~\lnm{Abramenko}\orcid{0000-0001-6466-4226}}%\sep
\author[addressref={aff3}]{\inits{V.B.}\fnm{V.B.}~\lnm{Yurchyshyn}\orcid{0000-0001-9982-2175}}%\sep

%%%%%%%%%%%%%%%%%%%%%%%%%%%%%%%%%%%%%%%%%%%%%%%%%%%
%% Affilations 
%% id shold be the same with \author addressref value.
\address[id=aff1]{Crimean Astrophysical Observatory, Russian Academy of Science, Nauchny, Bakhchisaray, 298409, Crimea, Russia}
\address[id=aff2]{Central (Pulkovo) Astronomical Observatory, Russian Academy of Science (GAO RAN), Pulkovskoye ch. 65,  Saint-Petersburg, 196140, Russia}
\address[id=aff3]{Big Bear Solar Observatory, New Jersey Institute of Technology, Big Bear City, CA 92314, USA}

%%%%%%%%%%%%%%%%%%%%%%%%%%%%%%%%%%%%%%%%%%%%%%%%%%%
%% Runningheads
%
\runningauthor{A.S.Kutsenko, V.I.Abramenko, V.B.Yurchyshyn}
\runningtitle{Contribution to the SMMF from Different Solar Regions}

%%%%%%%%%%%%%%%%%%%%%%%%%%%%%%%%%%%%%%%%%%%%%%%%%%% 
%%% Abstract 
\begin{abstract}

Seven-year long seeing-free observations of solar magnetic fields with the \textit{Helioseismic and Magnetic Imager}  (HMI) on board the \textit{Solar Dynamics Observatory} (SDO) were used to study the sources of the solar mean magnetic field, SMMF, defined as the net line-of-sight magnetic flux divided over the solar disk area. To evaluate the contribution of different regions to the SMMF, we separated all the pixels of each SDO/HMI magnetogram into three subsets: weak ($B$\textsuperscript{W}), intermediate ($B$\textsuperscript{I}), and strong ($B$\textsuperscript{S}) fields. The $B$\textsuperscript{W} component represents areas with magnetic flux densities below the chosen threshold; the $B$\textsuperscript{I} component is mainly represented by network fields, remains of decayed active regions (ARs), and ephemeral regions.  The $B$\textsuperscript{S} component consists of magnetic elements in ARs. To derive the contribution of a subset to the total SMMF, the linear regression coefficients between the corresponding component and the SMMF were calculated. We found that: i) when the threshold level of 30 Mx~cm\textsuperscript{-2} is applied,  the $B$\textsuperscript{I} and $B$\textsuperscript{S} components together contribute from 65\% to 95\% of the SMMF, while the fraction of the occupied area varies in a range of 2--6\% of the disk area; ii) as the threshold magnitude is lowered to 6 Mx~cm\textsuperscript{-2}, the contribution from $B$\textsuperscript{I}+$B$\textsuperscript{S} grows to 98\%, and the fraction of the occupied area  reaches the value of about 40\% of the solar disk. In summary, we found that regardless of the threshold level, only a small part of the solar disk area contributes to the SMMF. This means that the photospheric magnetic structure is an intermittent, inherently porous medium, resembling a percolation cluster. These findings suggest that the long-standing concept that continuous vast unipolar areas on the solar surface are the source of the SMMF may need to be reconsidered. 

\end{abstract}

%%%%%%%%%%%%%%%%%%%%%%%%%%%%%%%%%%%%%%%%%%%%%%%%%%%
%% Keywords
%

\end{opening}
%-------------------------------------------------

%%%%%%%%%%%%%%%%%%%%%%%%%%%%%%%%%%%%%%%%%%%%%%%%%%%
%% Sections Introduction
%
 \section{Introduction}
	 \label{sec-Introduction} 

The solar mean magnetic field (SMMF) is defined as a mean line-of-sight (LOS) magnetic field calculated over the entire visible solar disk \citep{Scherrer1977, Garcia1999}. Since most of the magnetic regions on the Sun are bipolar the SMMF has very low amplitude ranging from about 0.15~G during solar activity minimum to 1~G during maximum \citep{Livingston1991, Garcia1999}. Thus, the magnitude of the SMMF represents the imbalance of the LOS magnetic flux density calculated over the entire solar disk.

%{\S}{\bf ---} \\

The SMMF is closely related to the interplanetary magnetic field (IMF, \cite{Severny1970, Scherrer1977}. The SMMF structure and polarity are reproduced by \textit{in situ} observations of the IMF near the Earth five to seven days later \citep{Bremer1996}.

%{\S}{\bf ---} \\

A commonly accepted point of view is that the SMMF magnitude is defined by vast areas on the solar surface that exhibit weak, large-scale magnetic fields \citep{Severny1971, Scherrer1977, Kotov1977, Haneychuk2003, Xiang2016}. 

%{\S}{\bf ---} \\

On the other hand, there are observational indications suggesting that the SMMF is related, at least partly, to the presence of active regions (ARs) on the solar disk. \cite{Kotov1977} analyzed contribution of the background and AR magnetic fields to the SMMF, and concluded that while the SMMF is predominantly defined by vast regions of weak field, ARs can significantly contribute to the SMMF during certain time intervals. \cite{Demidov2011} also reported that the influence of ARs on the SMMF can be rather essential. Numerical simulations of the SMMF \citep{Sheeley1985}, that took into account only ARs on the solar disk as a source of magnetic flux, showed a good correlation with the observed SMMF. Indirect evidence of the relationship between the SMMF and ARs can also be inferred from sunspots statistics measured on time scales of a solar cycle \citep[\textit{e.g.}][]{Xiang2016}: a long-period (years) component of the SMMF varies almost in phase with the total area of sunspots. Both the SMMF and the total area of ARs reveal nearly 27 day variations caused by solar differential rotation \citep[\textit{e.g.}][]{Boberg2002, Haneychuk2003}.

%{\S}{\bf ---} \\

The current state of the research topic and the availability of high quality full disk magnetic field data motivated us to further explore the role of ARs in the SMMF formation.

%{\S}{\bf ---} \\

During the last couple of decades an impressive progress in solar instrumentation was made, which substantially advanced our understanding of solar magnetism in general and quiet-Sun magnetic fields in particular. Thus, analysis of \textit{Hinode}/SOT magnetograph data demonstrated the ubiquitous presence of mixed polarity magnetic concentrations along with unipolar isolated magnetic elements distributed all over the solar disk \citep{Ishikawa2010, Ishikawa2011, Lites2008}, and separated by vast zones of noise (see Figure 1 in \cite{Ishikawa2011}). The 1.6 m  \textit{New Solar Telescope} at the Big Bear Solar Observatory (NST/BBSO) allowed us to estimate the minimal observed size of magnetic flux tubes, which turned out to be less than the diffraction limit (77 km) of the NST \citep{Abramenko2010}. The magnetic field structure in areas outside ARs was found to possess  multifractal properties \citep{Abramenko2013}. All this encourages us to reexamine the long-standing paradigm that vast areas of weak magnetic fields outside ARs are essential contributors to the observed SMMF.

%{\S}{\bf ---} \\

Recently, we showed \citep{Kutsenko2016} that the SMMF can be calculated as an average of the magnetic flux density from LOS magnetograms provided by the \textit{Helioseismic and Magnetic Imager} \citep[HMI,][]{Schou2012} on board the \textit{Solar Dynamics Observatory} (SDO, \cite{Scherrer2012, Liu2012}). In the present article, we intend to determine the contribution to the SMMF made by different magnetic areas. We separated all pixels of SDO/HMI magnetograms into three subsets: i) noise and/or quiet-Sun pixels (the weakest magnetic areas), ii) AR areas (strongest magnetic areas), and iii) intermediate areas (predominantly remains of ARs, ephemeral regions, network fields). To evaluate their contributions, we determined a fraction of the magnetic flux from each subset in the total SMMF. The total area occupied by each subset was also scrutinized. The seven-year time interval starting April, 2010 was analyzed. 

%%%%%%%%%%%%%%%%%%%%%%%%%%%%%%%%%%%%%%%%%%%%%%%%%%%
%% Section Observables and methods

 \section{Observations and Data Analysis}
	 \label{sec-Observ} 
  \subsection{The SDO/HMI Data Set}
  \label{subs-Obs1}	 
  
Recently \cite{Kutsenko2016} showed that SDO/HMI magnetograms can be successfully used to calculate the SMMF by averaging the magnetic flux density in a full-disk magnetogram. A comparison to the patrol SMMF measurements performed at the Wilcox Solar Observatory (WSO, \cite{Scherrer1977}) showed that the conversion factor between the SDO/HMI and WSO SMMFs is very close to unity ($B$\textsuperscript{HMI} = (0.99$\pm$0.02)$B$\textsuperscript{WSO}) and the correlation coefficient equals 0.86. In the present study, we used one LOS magnetogram per day from the hmi.M\underline{ }720s series covering a seven-year interval from April 2010 to November 2016.

%{\S}{\bf ---} \\

\begin{figure}    %%%%%%%%%%%%%%%%%% FIGURE 1 
  	\centerline{\includegraphics[width=0.5\textwidth,clip=]{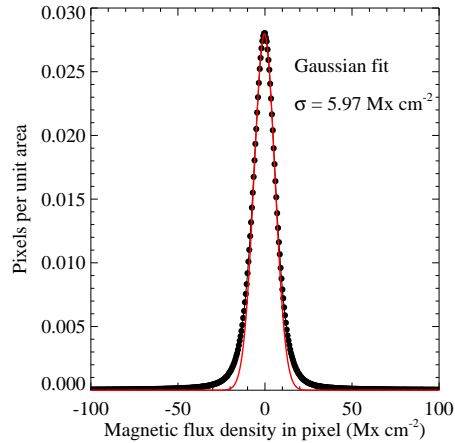}
  	}
  	\caption{A typical distribution function for the magnetic flux density in a SDO/HMI magnetogram (black dots). A Gaussian fit to the core of the distribution is shown in red. The standard deviation of the Gaussian is 5.97 Mx~cm\textsuperscript{-2}, while the offset of the fit curve is -0.27 Mx~cm\textsuperscript{-2}
  	}
  	\label{Fig1}
  \end{figure}

The instrument produces a 4096$\times$4096 pixel image with a pixel size 0.5" and a spatial resolution of 1" \citep{Scherrer2012, Schou2012, Liu2012}. A typical distribution function of the magnetic flux density in a hmi.M\underline{ }720s magnetogram is shown in Figure~\ref{Fig1}. Following \cite{Hagenaar2001} and \cite{Liu2012}, we approximated the core of the distribution function of each magnetogram by a Gaussian fit $F(B) = F\textsuperscript{MAX}$exp$(-(B-$offset$)\textsuperscript{2}/2\sigma\textsuperscript{2})$

%{\S}{\bf ---} \\

The noise level of magnetograms varies slightly around a mean value $\sigma$=5.9 Mx~cm\textsuperscript{-2} that is consistent with the result by \cite{Liu2012} who obtained $\sigma$=6.3 Mx~cm\textsuperscript{-2}. The noise level is determined by the instrumental noise as well as by the quiet-Sun fields.

%{\S}{\bf ---} \\

Following \cite{Hagenaar2001}, we accepted 5$\sigma$=30 Mx~cm\textsuperscript{-2} as the threshold value for discriminating between true solar magnetic elements and the artificial increase of magnetic flux density that might be caused by instrumental noise. Interestingly, the Gaussian fits to the distribution function of flux densities are typically shifted toward negative values by approximately 0.20 Mx~cm\textsuperscript{-2}. This differs from \cite{Liu2012}, who reported a positive offset of 0.07 Mx~cm\textsuperscript{-2}.

\subsection{Detection of Magnetic Flux Concentrations}
\label{subs-Obs2}	 
  
\begin{figure}    %%%%%%%%%%%%%%%%%% FIGURE 2 
  	\centerline{\includegraphics[width=1.0\textwidth,clip=]{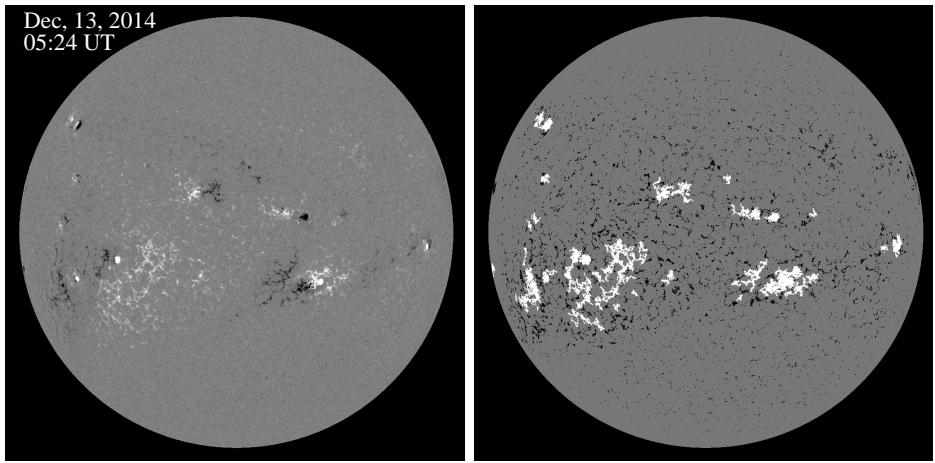}
  	}
  	\caption{
  		SDO/HMI full-disk 720s magnetogram on 13 December, 2014 (left panel) and the corresponding mask (right panel). The MFCs, which are attributed to ARs, are shown in white while other MFCs are shown in black. Quiet-Sun areas are shown in gray. The magnetogram is scaled from 500 Mx~cm\textsuperscript{-2} (white) to -500 Mx~cm\textsuperscript{-2} (black).
  	}
  	\label{Fig2}
  \end{figure}

A mask was built for each magnetogram to detect magnetic flux concentrations (MFCs). We consider a MFC as a group of touching pixels that comprises a connected magnetic region in the photosphere. The mask was created by thresholding a magnetogram with the value of 5$\sigma$. If the absolute magnetic flux density in a pixel exceeds the threshold, the correspondent pixel in the mask is set to unity. At the next step of the analysis a region-growing technique similar to that used by \cite{Benkhalil2006} was applied. A unity pixel nearest to the first element of the mask is selected as a ``seed" pixel.  The seed pixel is treated as the initial element of an MFC: pixels of the expected MFC will be clustered around it. This seed pixel is labeled uniquely and the four neighboring four pixels (located to the left, right, bottom, and top) are scanned. If any of the four pixels is also unity the pixel coordinates are stored and it is labeled as the initial seed pixel and the procedure is repeated. In case there are no new pixels the algorithm considers the cluster formation complete and proceeds to search for a next unity pixel. We analyzed only those MFCs with an area of three or more squared pixels and located within a circle of size 0.95 of the solar radius. As a result we obtained a map of MFCs on the photosphere. Each isolated MFC has a unique label and all its pixel coordinates are stored. The total flux of an MFC was then calculated as the sum of absolute values of the magnetic flux density in pixels multiplied by a pixel area. Note that groups of weak-field pixels (below the threshold) completely surrounded by MFCs remain uncaptured by these surrounding MFCs and, consequently, their magnetic flux does not contribute to the total magnetic flux of these MFCs. An example of such weak-field groups of pixels is shown in Figure~\ref{Fig2}, right panel.

%{\S}{\bf ---} \\

We first detected strong-flux MFCs, \textit{i.e.} those that belong to active regions (ARs). To identify an AR-type MFC, we used the definition by \cite{Hagenaar2003}: a MFC is attributed to the AR-type if its total unsigned flux exceeds 3$\times$10\textsuperscript{20} Mx. Note that an AR-type MFC may occupy an extended area and include mixed polarity magnetic fields. An example of the mask is shown in Figure~\ref{Fig2}, where the AR-type MFCs are shown in white. The contribution from all AR-type MFCs into the SMMF will be denoted as the $B$\textsuperscript{S} component. 

%{\S}{\bf ---} \\

The rest of the detected MFCs we considered as intermediate-type MFCs (Figure~\ref{Fig2}, black). Their contribution into the SMMF is denoted as the $B$\textsuperscript{I} component. 

%{\S}{\bf ---} \\

Finally, all pixels with the absolute flux density below the 5$\sigma$=30 Mx~cm\textsuperscript{-2} threshold represent the weakest field areas and their contribution to the SMMF is denoted here as the $B$\textsuperscript{W} component.

%{\S}{\bf ---} \\

Note that, with certain exceptions, the pixels once attributed to intermediate- or strong-flux regions (at the stage of mask building where the pixel separation by threshold is performed) cannot be reidentified as a weak-field pixels (at the stage of MFC flux calculation where MFCs are only identified as intermediate- or strong-flux regions) and \textit{vice versa}. The exceptions are pixels that compose a MFC of an area of one or two pixels as mentioned above although their influence is negligible. Note also that the noise may cause fragmentation of a MFC that belongs to a real single magnetic region. In this case the algorithm may fail to correctly identify the type of a region, nevertheless it affects only the $B$\textsuperscript{I} and $B$\textsuperscript{S} components.

%{\S}{\bf ---} \\

To estimate the contribution of each of the three types separately, we calculated the total signed magnetic flux over each of the AR-type and intermediate MFCs, and all weakest/noise pixels. The obtained fluxes were normalized by the total area of the solar disk so that the obtained values mimic the SMMF that would have been produced by a given MFC assuming that there are no other sources of magnetic flux on the disk. Naturally, the sum of these MFC fluxes gives the total SMMF. Note that the SMMF components are measured in Mx~cm\textsuperscript{-2} since they are obtained by dividing magnetic flux by the solar disk area. 

%{\S}{\bf ---} \\

A comparison of magnetograms and the corresponding masks (see Figure~\ref{Fig2}) allows us to infer what kind of MFCs are associated with the intermediate-flux component. On one hand, analysis of the mask pattern shows that the intermediate-flux MFCs form hexagonal patterns and, consequently, they are predominantly represented by network fields along the boundaries of supergranules. On the other hand, the weakest detected MFC consists of three pixels of 30 Mx~cm\textsuperscript{-2} and yields approximately 10\textsuperscript{17} Mx of magnetic flux. Hence we can conclude that the intermediate-flux component $B$\textsuperscript{I} is formed as a combination of: i) network magnetic elements with fluxes of 10\textsuperscript{17}-10\textsuperscript{19} Mx, ii) ephemeral regions with magnetic flux of 10\textsuperscript{18}-3$\times$10\textsuperscript{20} Mx \citep{Karak2016}, and iii) remains  of decayed ARs \citep{Hagenaar2003}.  
  
%%%%%%%%%%%%%%%%%%%%%%%%%%%%%%%%%%%%%%%%%%%%%%%%%%%
%% Section Results
 
 \section{Results}
	 \label{sec-Results}

\begin{figure}    %%%%%%%%%%%%%%%%%% FIGURE 3 
  	\centerline{\includegraphics[width=1.0\textwidth,clip=]{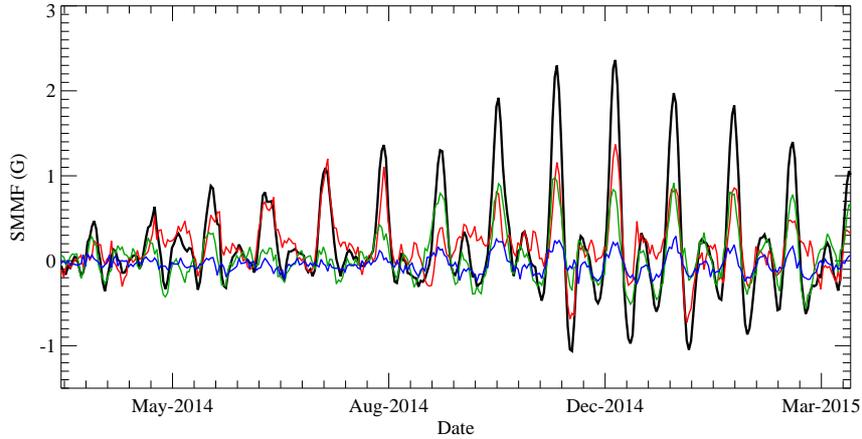}
  	}
  	\caption{One-year temporal variations of the components of the SMMF (see text). The weak, intermediate, and strong-flux components are shown in blue, green, and red colors, respectively. The total SMMF is overplotted in black.
  	}
  	\label{Fig3}
  \end{figure}

Following the above approach we calculated the three components of the SMMF and the total SMMF from each of the 2392 magnetograms in the seven-year long dataset. A sample of their time variations is shown in Figure~\ref{Fig3}. The weak, intermediate, and AR-flux components are plotted in blue, green, and red colors, respectively. The total SMMF is shown in black. The plot covers a one-year interval from April 2014 to March 2015. One can see that the weak-flux component, $B$\textsuperscript{W}, slightly undulates around the zero level, while the intermediate, $B$\textsuperscript{I}, and the strong-flux, $B$\textsuperscript{S}, components exhibit relatively intense, often similar in amplitude variations. Also there are intervals where the total SMMF is determined predominantly by the strong-flux component as, for example, during June-August 2014.

%{\S}{\bf ---} \\

\begin{figure}    %%%%%%%%%%%%%%%%%% FIGURE 4 
	\centerline{\includegraphics[width=1.0\textwidth,clip=]{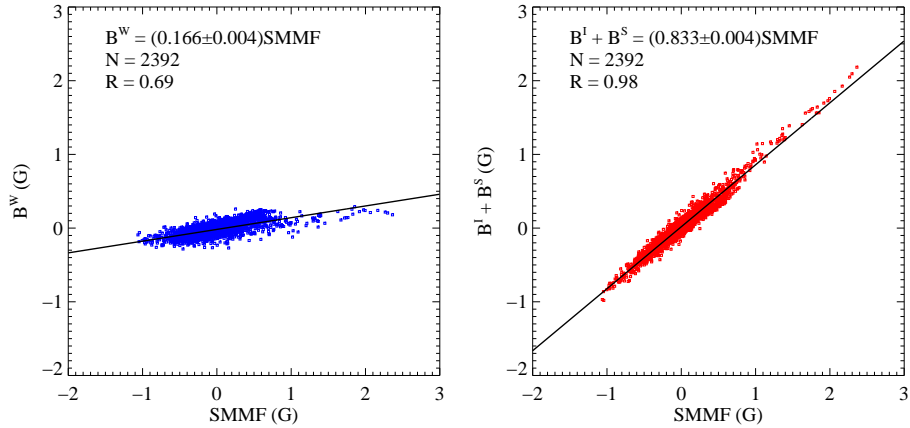}
	}
	\caption{Comparison of the total SMMF with the weak-flux component (left panel) and with the sum of the intermediate-flux and the strong-flux components (right panel). Linear regression coefficient, number of points, and the Pearson's correlation coefficient are shown for each plot.
	}
	\label{Fig4}
\end{figure}

To estimate the contribution of a given component to the total SMMF during the entire period under study, we calculated linear regression coefficients between the component and the total SMMF (Figure~\ref{Fig4}), which shows that the $B$\textsuperscript{W} component contributes on average only 17\% to the total SMMF, while the sum $B$\textsuperscript{I}+$B$\textsuperscript{S} contributes the remaining 83\%. The Pearson's correlation coefficient between $B$\textsuperscript{I}+$B$\textsuperscript{S} and the total SMMF equals 0.98 whereas the dependence between the SMMF and $B$\textsuperscript{W} is weaker: the Pearson's R in this case equals 0.69. Note that $B$\textsuperscript{I} and $B$\textsuperscript{S} separately contribute 49\% and 34\%, respectively.

\begin{figure}    %%%%%%%%%%%%%%%%%% FIGURE 5 
  	\centerline{\includegraphics[width=1.0\textwidth,clip=]{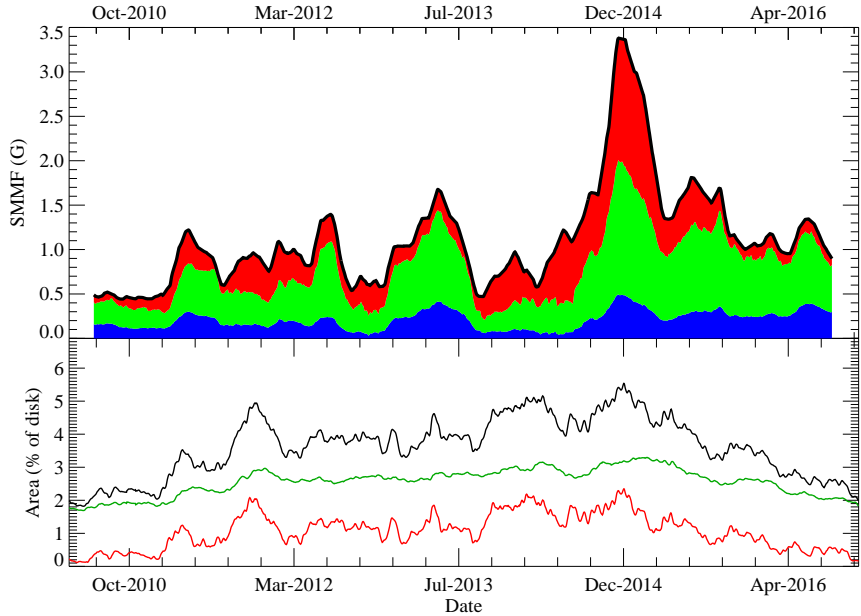}
  	}
  	\caption{Top -- The peak-to-peak amplitude of the observed SMMF (thick black curve, see text). Blue, green, and red colored horizontal ribbons under the black curve represent the values of $B$\textsuperscript{W},  $B$\textsuperscript{I}, and $B$\textsuperscript{S} components, respectively. The variation of a fraction of a certain component in the total SMMF is proportional to the width (in the vertical direction) of the corresponding ribbon. Bottom -- The variation of the total area of the sum $B$\textsuperscript{I}+$B$\textsuperscript{S} in terms of percentage of the disk area (black), total area of the intermediate-flux MFCs, $B$\textsuperscript{I} (green), and the total area of the strong-flux MFCs, $B$\textsuperscript{S} (red). All the area curves are averaged over the 27-day interval.
  	}
  	\label{Fig5}
  \end{figure}

%{\S}{\bf ---} \\
We also examined temporal variations of the contributions with a solar cycle. We chose a 150-day-long window initially centered on July, 15, 2010. The SMMF and its three component values related to this temporal interval were extracted from the data series and linear regression coefficients between the total SMMF and the $B$\textsuperscript{W}, $B$\textsuperscript{I}, and $B$\textsuperscript{S} components were calculated. Then the window center was shifted one day forward and the procedure repeated until the window center reached 01 September, 2016. The results are the temporal variation plots with 2241 data points shown in the top panel of Figure~\ref{Fig5}. The window width was selected to cover at least several solar rotations to suppress 27-day variations. On the other hand, too long window might cause loss of information on relatively fast temporal variations.

The thick black curve in Figure~\ref{Fig5} shows the peak-to-peak amplitude of the SMMF. To calculate the peak-to-peak amplitude we applied a procedure suggested by \cite{Sheeley2015}. For each point on the plot, we chose a 28-day interval coaligned with the center of the 150-day-long window described above. The amplitude was calculated as an absolute difference of the maximum value of the SMMF minus the minimum value of the SMMF in the interval. The center of the interval was shifted synchronously with the 150-day-long window center. At a final step, the curve was smoothed by a 28-day-long sliding average. The procedure performs demodulation of the SMMF amplitude caused by the solar rotation.

The blue, green, and red colored horizontal ribbons under the black curve in Figure~\ref{Fig5} represent the values of the $B$\textsuperscript{W},  $B$\textsuperscript{I}, and $B$\textsuperscript{S} components, respectively. The contribution of each component is proportional to the width (in the vertical direction) of the corresponding colored ribbon. One can see, for example, that in 2016 the strong-flux component (the red ribbon in Figure~\ref{Fig5}, top panel), $B$\textsuperscript{S}, is a rather weak contributor.

%{\S}{\bf ---} \\

Figure~\ref{Fig5} (top panel) shows an interesting tendency in the $B$\textsuperscript{W} behavior. In 2010, which is the year that followed the minimum of solar activity, due to relatively low $B$\textsuperscript{I} and $B$\textsuperscript{S} values,  the weak-flux component, $B$\textsuperscript{W}, contributes up to one third to the SMMF. A tendency of a gradual growth of the fraction of $B$\textsuperscript{W} is observed after December, 2014, \textit{i.e.} during the declining phase of solar cycle 24.

%{\S}{\bf ---} \\

In general, during the seven-year observational interval, the sum $B$\textsuperscript{I}+$B$\textsuperscript{S} (black curve in the top panel of Figure~\ref{Fig5}) constituted between 65\% to 95\% of the SMMF.

%{\S}{\bf ---} \\

The highest values of the SMMF during cycle 24 were observed in November-December 2014 (Figure~\ref{Fig3}) reaching a magnitude of up to 3 G \citep{Kutsenko2016}. \cite{Sheeley2015} thoroughly analyzed consequent Carrington maps of HMI photospheric field during CR2155-2160 to reveal the reasons for this increase. They concluded that the increase was caused by ``the systematic emergence of flux in active regions whose longitudinal distribution greatly increased the Sun's dipole moment". We also made a brief analysis of this event. The active region NOAA 12192, the largest naked-eye sunspot group since 1990 \citep{Sheeley2015}, culminated on October 23, 2014 and started to decay (CR2156).  In addition, during the next rotation (CR2157) a new active region emerged among the dispersed following part of the former AR12192 \citep[see Figure 2 in][]{Sheeley2015} that probably led to cancellation and a decrease of some portion of the following part of negative-polarity magnetic flux. Due to the separation of the following and leading parts, on 13 December, 2014 (CR2158), the positive-polarity remains of AR12192 dominated the east-south quarter of the visible solar disk (Figure~\ref{Fig2}, left panel). The mask of the HMI magnetogram (Figure~\ref{Fig2}, right panel) indicates that this dispersed region was identified by our algorithm as a strong-flux MFC instead of an intermediate-flux one. This mistaken identification can be explained by the large area occupied by the MFC: the algorithm computes total magnetic flux over the entire MFC area rather than averaged over the MFC magnetic flux density. Although the hexagonal patterns imply that the region is practically decayed, the field strength in its pixels reaches values up to 700 Mx~cm\textsuperscript{-2}. Thus, the positive uncompensated magnetic flux of this region produced a rise in the SMMF. \cite{Sheeley2015} remarked that this event is not unique: such an increase of the SMMF was observed during previous cycles and it usually marks the onset of the declining phase of the cycle.

%{\S}{\bf ---} \\

Time variations of the area fraction corresponding to the $B$\textsuperscript{I} and $B$\textsuperscript{S} components are shown in Figure~\ref{Fig5}, bottom panel. The fraction of the total area of the MFCs that form intermediate- and strong-flux components, p($B$\textsuperscript{I}+$B$\textsuperscript{S}), is shown in the same panel in black, and it does not exceed 6\%. Note that there is not a perfect agreement between the area and the fraction of a given component (Figure~\ref{Fig5}, top panel). The area of the $B$\textsuperscript{I} component smoothly increases during the rising phase of the cycle and smoothly decreases during the declining phase. At the same time the absolute value of $B$\textsuperscript{I} fluctuates significantly during the same interval. Both the $B$\textsuperscript{I} and the $B$\textsuperscript{S} components exhibit variations that are not in phase with the variations of the corresponding areas (see Section~\ref{sec-Conclusions} for discussion on the relationship between the area and the net magnetic flux of MFCs).

%{\S}{\bf ---} \\

\begin{figure}    %%%%%%%%%%%%%%%%%% FIGURE 6 
	\centerline{\includegraphics[width=1\textwidth,clip=]{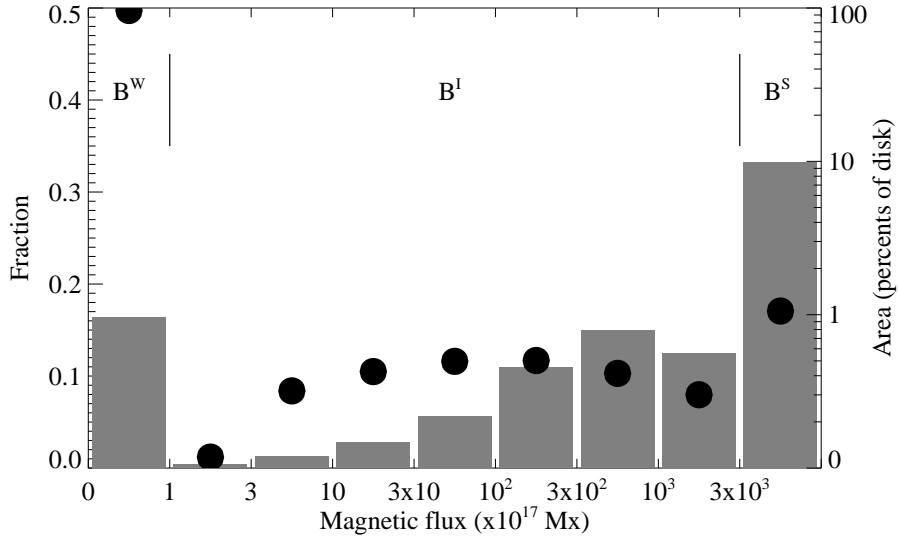}
	}
	\caption{Bars -- fraction in the SMMF from the MFCs with the flux in a given bin, averaged over the seven-year interval data (left axis). Circles -- total area of MFCs in bins (percentage of the disk area, right axis).
	}
	\label{Fig6}
\end{figure}

To reveal what kind of MFCs contribute the most to the $B$\textsuperscript{I} component, we binned the absolute value of the flux of the intermediate-flux MFCs and calculated the contributions from each bin (see Figure~\ref{Fig6}, gray bars). For each bin, we then determined the linear regression coefficients (Figure~\ref{Fig6}) which show that when considering only the intermediate-flux MFCs, maximal contribution to the SMMF comes from the MFCs with magnetic flux of 1-30$\times$10\textsuperscript{19} Mx and they are predominantly associated with vast remains of decaying ARs. Note that a smooth maximum of the area distribution of the intermediate-flux MFCs (circles in Figure~\ref{Fig6}) does not coincide with the maximum of the fraction, being shifted toward the smaller fluxes.

%{\S}{\bf ---} \\

%%%%%%%%%%%%%%%%%%%%%%%%%%%%%%%%%%%%%%%%%%%%%%%%%%%
%% Section Results

\section{Possible Influence of the Choice of Threshold}
\label{sec-Threshold}

\begin{figure}    %%%%%%%%%%%%%%%%%% FIGURE 7 
  	\centerline{\includegraphics[width=1\textwidth,clip=]{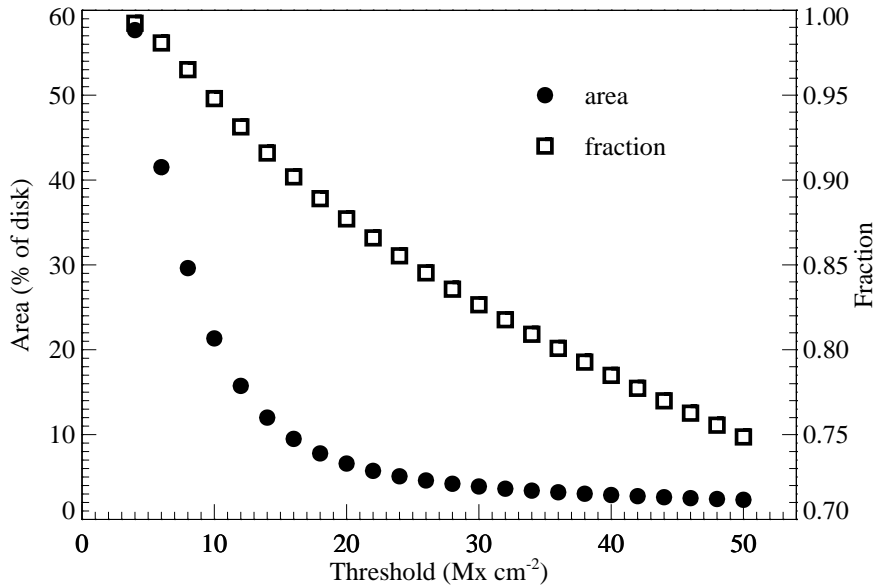}
  	}
  	\caption{Area occupied by intermediate-flux MFCs and strong-flux MFCs, p($B$\textsuperscript{I}+$B$\textsuperscript{S}), versus the threshold value (circles). Fraction in the SMMF of $B$\textsuperscript{I}+$B$\textsuperscript{S} MFCs (squares). Data were averaged over the seven-year time interval.
  	}
  	\label{Fig7}
  \end{figure}

The above analysis was performed with the threshold of 30 Mx~cm\textsuperscript{-2}. In order to understand how the choice of the threshold affects the result we repeated the calculations for a set of thresholds ranging from 4 to 50 Mx~cm\textsuperscript{-2} with a sampling of 2 Mx~cm\textsuperscript{-2}. The fraction of the total area of the sum $B$\textsuperscript{I}+$B$\textsuperscript{S} averaged over the seven-year interval, as well as their joint contribution to the SMMF, are plotted versus the threshold in Figure~\ref{Fig7}. Changing the threshold from, say, 50 to 12 Mx~cm\textsuperscript{-2} (twice the noise level) results in a change in the total area from 4\% to 16\%. The total $B$\textsuperscript{I}+$B$\textsuperscript{S} area is still small compared to the area of weak-flux pixels and the joint contribution of $B$\textsuperscript{I}+$B$\textsuperscript{S} (0.75 at the 50 Mx~cm\textsuperscript{-2} threshold and 0.93 at the 12 Mx~cm\textsuperscript{-2} threshold) still dominates in the SMMF. Further decrease of the threshold results in a rapid increase of all parameters by up to 41\% and 0.98 at the noise level 6 Mx~cm\textsuperscript{-2}. The experiment demonstrates that only a small portion of the solar disk area determines the SMMF value and the choice of the threshold does not critically affect the results.

%{\S}{\bf ---} \\

Strictly speaking, a MFC can transit from the AR-type subset to the intermediate-flux subset and \textit{vice versa} due to decay, fragmentation, emergence of new flux, \textit{etc.}, as an AR passes across the solar disk. This may lead to rapid decrease of $B$\textsuperscript{S} and increase of $B$\textsuperscript{I}. Also a MFC which belongs to a fragmented part of an AR can be identified as an intermediate-flux MFC. All this might affect the time profiles of the components.  The damage can be estimated, in particular, by lowering the threshold of 3$\times$10\textsuperscript{20} Mx determining the AR-type MFCs. Thus we checked all the intermediate-flux MFCs with magnetic flux greater than 1$\times$10\textsuperscript{20} Mx (three times less than the AR threshold). If there was an AR-type MFC in the vicinity of the intermediate-flux MFC (approximately at a distance of a single supergranule, 30 Mm) then this intermediate-flux MFC was identified as an AR-type MFC belonging to the closest AR. This procedure added dispersed isolated AR` remains to the AR-type MFCs. As a result the averaged contribution of the ARs to the SMMF increased from 34\% to 43\%, while the contribution of the $B$\textsuperscript{I} component decreased from 48\% to 41\%. The changes are noticeable but not significant. This implies that both these sources contribute almost equally to the SMMF regardless of the identification technique.

%{\S}{\bf ---} \\

%%%%%%%%%%%%%%%%%%%%%%%%%%%%%%%%%%%%%%%%%%%%%%%%%%%
%% Section Conclusion

 \section{Conclusions and Discussion}
	  \label{sec-Conclusions}
		  
An opportunity for seven-year long uninterrupted seeing-free observations of the solar magnetic fields with the SDO/HMI instrument was used here to study the sources of the solar mean magnetic field, SMMF, defined as the net line-of-sight magnetic flux divided by the solar disk area.

%{\S}{\bf ---} \\
 
To evaluate the contribution of different magnetic regions on the solar disk to the SMMF, we separated all the pixels of each SDO/HMI magnetogram into three subsets. Each subset predominantly includes a certain magnetic-activity region at the photosphere comprising weak ($B$\textsuperscript{W}), intermediate ($B$\textsuperscript{I}), or strong ($B$\textsuperscript{S}) fields. The $B$\textsuperscript{W} component represents areas with magnetic flux densities below the chosen threshold; the $B$\textsuperscript{I} component is represented mainly by network fields, remains of decayed ARs, and ephemeral regions. The last component, $B$\textsuperscript{S}, consists of magnetic elements in ARs. To measure the contribution of a certain MFC subset to the total SMMF, the linear regression coefficients between the corresponding component and the SMMF were calculated.

%{\S}{\bf ---} \\

We found the following:

i) With the threshold level of 30 Mx~cm\textsuperscript{-2} (five standard deviations of the HMI noise level), the intermediate- and strong-flux components, $B$\textsuperscript{I} and $B$\textsuperscript{S}, together contribute from 65\% to 95\% of the SMMF. At the same time, the fraction of the area, p($B$\textsuperscript{I}+$B$\textsuperscript{S}), varies in the range of 2-6\% of the disk area.

ii) As the threshold magnitude is lowered to one standard deviation of the HMI noise level, 6 Mx~cm\textsuperscript{-2}, the contribution from the $B$\textsuperscript{I}+$B$\textsuperscript{S} component grows to 98\%, and the fraction of the area, p($B$\textsuperscript{I}+$B$\textsuperscript{S}), reaches the value of about 40\% of the solar disk.

iii) The intermediate-flux component, $B$\textsuperscript{I}, and the strong-flux component,  $B$\textsuperscript{S}, contribute almost equally into the SMMF regardless of the identification procedure details.

%{\S}{\bf ---} \\

According to the pioneering studies of the SMMF \citep{Severny1971, Scherrer1977, Kotov1977}, the SMMF was thought to be predominantly determined by vast areas of quiet-Sun regions covering more than 80\% of the solar surface. Such an overestimation of the quiet-Sun areas can be explained by shortcomings of the data available at that time, in particular, by the low spatial resolution of full-disk magnetograms. Indeed, \cite{Severny1971} pointed out that ``resolution plays a most important role, because the effect of averaging at low resolution seriously distorts the pattern obtained with the magnetograph and can produce fictitious large-scale unipolar regions". Thus, strong magnetic fields of relatively compact and isolated regions of AR remains become ``smeared" over the integration area determined by the spatial resolution of the instrument. Since decayed ARs exhibit neither sunspots nor pores in continuum so there are no means to separate whether the magnetic field element originated in a small strong-flux region or a vast region of weak fields.

%{\S}{\bf ---} \\

As we showed above, MFCs of magnetic flux of 100-3000$\times$10\textsuperscript{17} Mx make the largest contribution to the intermediate-field component, $B$\textsuperscript{I}. Meanwhile the peak of the total area distribution of the intermediate-flux subsets is shifted towards lower magnetic fluxes. \cite{MunozJaramillo2015} argued that, at least down to fluxes of 10\textsuperscript{19} Mx, the magnetic flux of a magnetic element is proportional to its area. We may then assume that, as a rough approximation, the same dependence extends to our lower limit of 10\textsuperscript{17} Mx. In this case, the MFCs with magnetic flux of 10-1000$\times$10\textsuperscript{17} Mx posses the majority of the total unsigned magnetic flux among the intermediate-flux MFC subset, although they do not provide the largest contribution to the net flux. \cite{Hagenaar2003} concluded that the distribution of such small magnetic regions can be accurately approximated by a sum of two exponential functions. The low-flux end of the flux distribution represents small ephemeral regions of flux \textless10\textsuperscript{19}; their emergence is almost independent on the cycle phase and these ephemeral regions emerge mostly as a small bipoles. Consequently, we can assume that their flux imbalance is small, \textit{i.e.} they do not contribute much to the SMMF. A slight increase of the emergence rate of these ephemeral regions is observed during cycle minimum \citep{Hagenaar2003, Karak2016} that confirms once again their rather weak influence on the SMMF: the SMMF exhibits in-phase variation with the solar cycle \citep[\textit{e.g.}][]{Xiang2016}. The high-flux (\textgreater10\textsuperscript{19} Mx) part of the distribution obtained by \cite{Hagenaar2003} is ``dominated by unipolar areas, in which the cycle modulation of the active regions that feed these areas is clearly reflected...". Thus, the intermediate-flux MFCs in our study that contribute the most to the SMMF are represented by unipolar remains of decayed ARs, which supports our conclusion based on the analysis of the map of these MFCs.

%{\S}{\bf ---} \\

The surprisingly high contribution of ARs to the SMMF may explain the rapid variations of the SMMF (minutes to hours) that are not related to solar rotation \citep[\textit{e.g.}][]{Demidov1995, Demidov2011}. Indeed, during certain intervals, when the contribution of ARs to the SMMF is high, short-period oscillations in the SMMF can be caused by magnetic flux changes inside ARs. Anyway, this topic requires further investigation.

%{\S}{\bf ---} \\

The weak-field component contribution to the SMMF decreases from 16\% to 2\% as the detection threshold is lowered from 30 to 6 Mx~cm\textsuperscript{-2}. This low contribution into the total net flux can be explained by the predominantly closed nature of weak intranetwork ``salt-and-pepper" magnetic features, see, \textit{e.g.}, the high-resolution observations findings \citep{Lites2008, Ishikawa2010, Abramenko2013}. At the same time, we show here that the weak-flux component, calculated with the threshold of 30 Mx~cm\textsuperscript{-2}, exhibits 27-day variations (see Figure~\ref{Fig3}) in phase with that of the strong-flux component. This implies that both variations might be caused by the same physical mechanisms, namely the high concentrations of magnetic flux at ``active longitudes" (see, \textit{e.g.}, \citealp{vanDrielGesztelyi2015}). We assume that this happens due to fragments of ARs less than 10\textsuperscript{17} Mx. Recall that it takes more than a year for magnetic flux injected near the equator to migrate to a pole \citep{Sheeley1986}. So the small unipolar remains can stay at near the same place for many solar rotations. 

%{\S}{\bf ---} \\

Regardless of the threshold level, only a small part of the solar disk area is covered by magnetic flux with density above the threshold value. This means that the photospheric magnetic structure is an inherently porous medium, resembling a percolation cluster. In other words, one can say that vast areas of low activity are intermittent with rather small zones of high activity -- a classical definition of an intermittent structure with fractal properties \citep[\textit{e.g.}][]{Frisch1995}. The multifractal behavior of various local patterns on the Sun (ARs, zones in the undisturbed photosphere, \textit{etc.}) are well demonstrated in the literature \citep[\textit{e.g.}][and references therein]{Lawrence1993, Abramenko2008, McAteer2016}. The property of multifractality of the entire solar disk has not been well explored so far. 

%{\S}{\bf ---} \\

As for the revealed signature of percolation, this finding seems to be a promising  venue to study solar magnetism, similar to other processes in the cosmic plasma \citep[see, \textit{e.g.},][]{Milovanov2001}. The first steps to apply the percolation theory in the solar field of research were already undertaken \citep[see, \textit{e.g.},][]{Wentzel1992, Pustilnik1999, Schatten2007}. The inherent properties of any percolation process -- power-law spectra, self-organization of large structures amid chaos, spontaneous transitions  into a critical state -- are of vital importance for our further understanding of the nature of solar magnetism.

%% Figure 
%
% \begin{figure} 
% \centerline{\includegraphics[width=0.5\textwidth,clip=]{<fig.eps>}}
% \caption{}%\label{fig:?}
% \end{figure}

%% Table
%
% \begin{table}
% \caption{}%\label{tbl:?}
% \begin{tabular}{}     
% \hline
% \multicolumn{2}{c}{<>}
% <data>
% \hline
% \end{tabular}
% \end{table}

%%%%%%%%%%%%%%%%%%%%%%%%%%%%%%%%%%%%%%%%%%%%%%%%%%%%%%%%%%%%%%%%%%%%%%%%%%%
%% Appendix
%
% \appendix   

%%%%%%%%%%%%%%%%%%%%%%%%%%%%%%%%%%%%%%%%%%%%%%%%%%%%%%%%%%%%%%%%%%%%%%%%%%%
%% Acknowledgements
%
 \begin{acks}
 	
	%{\S}{\bf --- Acks} \\
We are grateful to the anonymous referees for their criticism and raised questions that helped us to improve the article. SDO is a mission for NASA's Living With a Star (LWS) program. The SDO/HMI data were provided by the Joint Science Operation Center (JSOC). The reported study was supported in part by the RFBR research projects 16-42-910493, 16-02-00221 A, 17-02-00049 A and the Presidium of the Russian Academy of Science Program 7. VYu acknowledges support from AFOSR FA9550-15-1-0322 and NSF AGS-1250818 grants.

	%{\S}{\bf --- Acks} \\

\end{acks}

\textbf{Disclosure of Potential Conflicts of Interest} The authors declare that they have no conflicts of interest.

%%% %%%%%%%%%%%%%%%%%%%%%%%%%%%%%%%%%%%%%%%%%%%%%%%%%%%%%%%%%%%
%% Bibliography
%
% Using BibTeX
%
%\bibliographystyle{spr-mp-sola}
%\bibliography{Kutsenko_Abramenko} 

\begin{thebibliography}{}

\bibitem[Abramenko(2013)]
	{Abramenko2013}
	Abramenko, V.I.: 2013, {\it Proc.  IAU Symp.} {\bf 294}, 289. \href{http://dx.doi.org/10.1017/S1743921313002652}{DOI} \href{http://adsabs.harvard.edu/abs/2013IAUS..294..289A}{ADS} 
	
\bibitem[Abramenko \emph{et al.}(2010)]
	{Abramenko2010}
	Abramenko, V., Yurchyshyn, V., Goode, P., Kilcik, A.: 2010, {\it Astrophys. J.} {\bf 725}, L101. \href{http://dx.doi.org/10.1088/2041-8205/725/1/L101}{DOI} \href{http://adsabs.harvard.edu/abs/2010ApJ...725L.101A}{ADS} 
	
\bibitem[Abramenko, Yurchyshyn, and Wang(2008)]
	{Abramenko2008}
	Abramenko, V., Yurchyshyn, V., Wang, H.: 2008, {\it Astrophys. J.} {\bf 681}, 1669-1676. \href{http://dx.doi.org/10.1086/588426}{DOI} \href{http://adsabs.harvard.edu/abs/2008ApJ...681.1669A}{ADS} 

\bibitem[Benkhalil \emph{et al.}(2006)]
	{Benkhalil2006}
	Benkhalil, A., Zharkova, V.V., Zharkov, S., Ipson, S.: 2006, {\it Solar Phys.} {\bf 235}, 87. \href{http://dx.doi.org/10.1007/s11207-006-0023-7}{DOI} \href{http://adsabs.harvard.edu/abs/2006SoPh..235...87B}{ADS} 

\bibitem[Boberg \emph{et al.}(2002)]
	{Boberg2002}
	Boberg, F., Lundstedt, H., Hoeksema, J.T., Scherrer, P.H., Liu, W.: 2002, {\it J.Geophys. Res. A} {\bf 107}, 1318. \href{http://dx.doi.org/10.1029/2001JA009195}{DOI} \href{http://adsabs.harvard.edu/abs/2002JGRA..107.1318B}{ADS} 

\bibitem[Bremer(1996)]
	{Bremer1996}
	Bremer, J.: 1996, {\it Ann. Geophys.} {\bf 39}, 713. \href{http://dx.doi.org/10.4401\%2Fag-4003}{DOI}

\bibitem[Demidov(1995)]
	{Demidov1995}
	Demidov, M.L.: 1995, {\it Solar Phys.} {\bf 159}, 23. \href{http://dx.doi.org/10.1007/BF00733028}{DOI} \href{http://adsabs.harvard.edu/abs/1995SoPh..159...23D}{ADS} 

\bibitem[Demidov(2011)]
	{Demidov2011}
	Demidov, M.: 2011, {\it Physics of Sun and Star Spots} {\bf 273}, 56. \href{http://dx.doi.org/10.1017/S1743921311015006}{DOI} \href{http://adsabs.harvard.edu/abs/2011IAUS..273...56D}{ADS}

\bibitem[Frisch(1995)]
	{Frisch1995}
	Frisch, U., Turbulence, The Legacy of A.N. Kolmogorov / U. Frisch —  Cambridge: Cambridge University Press, 1995. — 296 p.

\bibitem[Garc{\'{\i}}a \emph{et al.}(1999)]
	{Garcia1999}
	Garc{\'{\i}}a, R.A., Boumier, P., Charra, J., Foglizzo, T., Gabriel, A.H., Grec, G., R{\'e}gulo, C., Robillot, J.M., Turck-Chi{\`e}ze, S., Ulrich, R.K.: 1999, {\it Astron. Astroph.} {\bf 346}, 626. \href{http://adsabs.harvard.edu/abs/1999A\%26A...346..626G}{ADS}

\bibitem[Hagenaar(2001)]
	{Hagenaar2001}
	Hagenaar, H.J.: 2001, {\it Astrophys. J.} {\bf 555}, 448. 
	\href{http://dx.doi.org/10.1086/321448}{DOI} \href{http://adsabs.harvard.edu/abs/2001ApJ...555..448H}{ADS} 


\bibitem[Hagenaar, Schrijver, and Title(2003)]
	{Hagenaar2003}
	Hagenaar, H.J., Schrijver, C.J., Title, A.M.: 2003, {\it Astrophys. J.} {\bf 584}, 1107.
	\href{http://dx.doi.org/10.1086/345792}{DOI} \href{http://adsabs.harvard.edu/abs/2003ApJ...584.1107H}{ADS} 

\bibitem[Haneychuk, Kotov, and Tsap(2003)]
	{Haneychuk2003}
	Haneychuk, V.I., Kotov, V.A., Tsap, T.T.: 2003, {\it Astron. Astroph.} {\bf 403}, 1115. \href{http://dx.doi.org/10.1051/0004-6361:20030426}{DOI} \href{http://adsabs.harvard.edu/abs/2003A\%26A...403.1115H}{ADS} 

\bibitem[Ishikawa and Tsuneta(2010)]
	{Ishikawa2010}
	Ishikawa, R., Tsuneta, S.: 2010, {\it Astrophys. J.} {\bf 718}, L171. 
	\href{http://dx.doi.org/10.1088/2041-8205/718/2/L171}{DOI} \href{http://adsabs.harvard.edu/abs/2010ApJ...718L.171I}{ADS} 

\bibitem[Ishikawa and Tsuneta(2011)]
	{Ishikawa2011}
	Ishikawa, R., Tsuneta, S.: 2011, {\it Astrophys. J.} {\bf 735}, 74. \href{http://dx.doi.org/10.1088/0004-637X/735/2/74}{DOI} \href{http://adsabs.harvard.edu/abs/2011ApJ...735...74IH}{ADS} 

\bibitem[Karak and Brandenburg(2016)]
	{Karak2016}
	Karak, B.B., Brandenburg, A.: 2016, {\it Astrophys. J.} {\bf 816}, 28. 
	\href{http://dx.doi.org/10.3847/0004-637X/816/1/28}{DOI} \href{http://adsabs.harvard.edu/abs/2016ApJ...816...28K}{ADS} 

\bibitem[Kotov, Stepanian, and Shcherbakova(1977)]
	{Kotov1977}
	Kotov, V.A., Stepanian, N.N., and Shcherbakova, Z.A.: 1977, {\it Izvestiya Ordena Trudovogo Krasnogo Znameni Krymskoj Astrofizicheskoj Observatorii} {\bf 56}, 75 (in Russian). \href{http://adsabs.harvard.edu/abs/1977IzKry..56...75K}{ADS} 

\bibitem[Kutsenko and Abramenko(2016)]
	{Kutsenko2016}
	Kutsenko, A.S., Abramenko, V.I.: 2016, {\it Solar Phys.} {\bf 291}, 1613. \href{http://dx.doi.org/10.1007/s11207-016-0940-z}{DOI} \href{http://adsabs.harvard.edu/abs/2016SoPh..291.1613K}{ADS} 

\bibitem[Lawrence, Ruzmaikin, and Cadavid(1993)]
	{Lawrence1993}
	Lawrence, J.K., Ruzmaikin, A.A., Cadavid, A.C.: 1993, {\it Astrophys. J.} {\bf 417}, 805.
	\href{http://dx.doi.org/10.1086/173360}{DOI} \href{http://adsabs.harvard.edu/abs/1993ApJ...417..805L}{ADS} 

\bibitem[Lites \emph{et al.}(2008)]
	{Lites2008}
	Lites, B.W., Kubo, M., Socas-Navarro, H., Berger, T., Frank, Z., Shine, R., Tarbell, T., Title, A., Ichimoto, K., Katsukawa, Y., Tsuneta, S., Suematsu, Y., Shimizu, T., Nagata, S.: 2008, {\it Astrophys. J.} {\bf 672}, 1237-1253. 	\href{http://dx.doi.org/10.1086/522922}{DOI} \href{http://adsabs.harvard.edu/abs/2008ApJ...672.1237L}{ADS} 
	
\bibitem[Liu \emph{et al.}(2012)]
	{Liu2012}
	Liu, Y., Hoeksema, J.T., Scherrer, P.H., Schou, J., Couvidat, S., Bush, R.I., Duvall, T.L., Hayashi, K., Sun, X., Zhao, X.: 2012, {\it Solar Phys.} {\bf 279}, 295. \href{http://dx.doi.org/10.1007/s11207-012-9976-x}{DOI} \href{http://adsabs.harvard.edu/abs/2012SoPh..279..295L}{ADS} 

\bibitem[Livingston \emph{et al.}(1991)]
	{Livingston1991}
	Livingston, W., Donnelly R.F., Grigoryev V., et al.: 1991, In: Cox A.N., Livingston W.C., Matthews M.S. (eds.) Solar Interior and Atmosphere. The University of Arizona Press 

\bibitem[McAteer \emph{et al.}(2016)]
	{McAteer2016}
	McAteer, R.T.J., Aschwanden, M.J., Dimitropoulou, M., Georgoulis, M.K., Pruessner, G., Morales, L., Ireland, J., Abramenko, V.: 2016, {\it Space Science Reviews} {\bf 198}, 217. \href{http://dx.doi.org/10.1007/s11214-015-0158-7}{DOI} \href{http://adsabs.harvard.edu/abs/2016SSRv..198..217M}{ADS} 

\bibitem[Milovanov \emph{et al.}(2001)]
	{Milovanov2001}
	Milovanov, A.V., Zelenyi, L.M., Zimbardo, G., Veltri, P.: 2001, {\it Journal of Geophysical Research} {\bf 106}, 6291.
	\href{http://dx.doi.org/10.1029/1999JA000446}{DOI} \href{http://adsabs.harvard.edu/abs/2001JGR...106.6291M}{ADS} 

\bibitem[Mu{\~n}oz-Jaramillo \emph{et al.}(2015)]
	{MunozJaramillo2015}
	Mu{\~n}oz-Jaramillo, A., Senkpeil, R.R., Windmueller, J.C., Amouzou, E.C., Longcope, D.W., Tlatov, A.G., Nagovitsyn, Y.A., Pevtsov, A.A., Chapman, G.A., Cookson, A.M., Yeates, A.R., Watson, F.T., Balmaceda, L.A., DeLuca, E.E., Martens, P.C.H.: 2015, {\it Astrophys. J.} {\bf 800}, 48. \href{http://dx.doi.org/10.1088/0004-637X/800/1/48}{DOI} \href{http://adsabs.harvard.edu/abs/2015ApJ...800...48M}{ADS}

\bibitem[Pustil'nik(1999)]
	{Pustilnik1999}
	Pustil'nik, L.A.: 1999, {\it Astronomical and Astrophysical Transactions} {\bf 18}, 227. \href{http://dx.doi.org/10.1080/10556799908203060}{DOI} \href{http://adsabs.harvard.edu/abs/1999A\%26AT...18..227P}{ADS}

\bibitem[Schatten(2007)]
	{Schatten2007}
	Schatten, K.H.: 2007, {\it The Astrophysical Journal Supplement Series} {\bf 169}, 137. \href{http://dx.doi.org/10.1086/510367}{DOI} \href{http://adsabs.harvard.edu/abs/2007ApJS..169..137S}{ADS} 

\bibitem[Scherrer \emph{et al.}(2012)]
	{Scherrer2012}
	Scherrer, P.H., Schou, J., Bush, R.I., Kosovichev, A.G., Bogart, R.S., Hoeksema, J.T., Liu, Y., Duvall, T.L., Zhao, J., Title, A.M., Schrijver, C.J., Tarbell, T.D., Tomczyk, S.: 2012, {\it Solar Phys.} {\bf 275}, 207. \href{http://dx.doi.org/10.1007/s11207-011-9834-2}{DOI} \href{http://adsabs.harvard.edu/abs/2012SoPh..275..207S}{ADS} 

\bibitem[Scherrer \emph{et al.}(1977)]
	{Scherrer1977}
	Scherrer, P.H., Wilcox, J.M., Kotov, V., Severnyi, A.B., and Howard, R.: 1977, {\it Solar Phys.} {\bf 52}, 3.
	\href{http://dx.doi.org/10.1007/BF00935783}{DOI} \href{http://adsabs.harvard.edu/abs/1977SoPh...52D...6S}{ADS} 

\bibitem[Schou \emph{et al.}(2012)]
	{Schou2012}
	Schou, J., Scherrer, P.H., Bush, R.I., Wachter, R., Couvidat, S., Rabello-Soares, M.C., Bogart, R.S., Hoeksema, J.T., Liu, Y., Duvall, T.L., Akin, D.J., Allard, B.A., Miles, J.W., Rairden, R., Shine, R.A., Tarbell, T.D., Title, A.M., Wolfson, C.J., Elmore, D.F., Norton, A.A., Tomczyk, S.: 2012, {\it Solar Phys.} {\bf 275}, 229. \href{http://dx.doi.org/10.1007/s11207-011-9842-2}{DOI} \href{http://adsabs.harvard.edu/abs/2012SoPh..275..229S}{ADS}

\bibitem[Severny \emph{et al.}(1970)]
	{Severny1970}
	Severny, A., Wilcox, J.M., Scherrer, P.H., Colburn, D.S.: 1970, {\it Solar Phys.} {\bf 15}, 3. \href{http://dx.doi.org/10.1007/BF00149468}{DOI} \href{http://adsabs.harvard.edu/abs/1970SoPh...15....3S}{ADS} 

\bibitem[Severny(1971)]
	{Severny1971}
	Severny, A.B.: 1971, {\it Quarterly Journal of the Royal Astronomical Society} {\bf 12}, 363. \href{http://adsabs.harvard.edu/abs/1971QJRAS..12..363S}{ADS}

\bibitem[Sheeley and DeVore(1986)]
	{Sheeley1986}
	Sheeley, N.R., Jr.~and DeVore, C.R.: 1986, {\it Solar Phys.} {\bf 103}, 203. \href{http://dx.doi.org/10.1007/BF00147824}{DOI} \href{http://adsabs.harvard.edu/abs/1986SoPh..103..203S}{ADS} 

\bibitem[Sheeley, DeVore, and Boris(1985)]
	{Sheeley1985}
	Sheeley, N.R., Jr., DeVore, C.R., Boris, J.P.: 1985, {\it Solar Phys.} {\bf 98}, 219. \href{http://dx.doi.org/10.1007/BF00152457}{DOI} \href{http://adsabs.harvard.edu/abs/1985SoPh...98..219S}{ADS} 

\bibitem[Sheeley and Wang(2015)]
	{Sheeley2015}
	Sheeley, N.R., Jr., Wang, Y.-M.: 2015, {\it Astrophys. J.} {\bf 809}, 113. 

\bibitem[van Driel-Gesztelyi and Green(2015)]
	{vanDrielGesztelyi2015}
	van Driel-Gesztelyi, L., Green, L.M.: 2015, {\it Living Reviews in Solar Physics} {\bf 12}, 1. 

\bibitem[Wentzel and Seiden(1992)]
	{Wentzel1992}
	Wentzel, D.G., Seiden, P.E.: 1992, {\it Astrophys. J.} {\bf 390}, 280. \href{http://dx.doi.org/10.1086/171278}{DOI} \href{http://adsabs.harvard.edu/abs/1992ApJ...390..280W}{ADS} 

\bibitem[Xiang and Qu(2016)]
	{Xiang2016}
	Xiang, N.B., Qu, Z.N.: 2016, {\it The Astronomical Journal} {\bf 151}, 76. \href{http://dx.doi.org/10.3847/0004-6256/151/3/76}{DOI} \href{http://adsabs.harvard.edu/abs/2016AJ....151...76X}{ADS}

 \end{thebibliography}
%\IfFileExists{\jobname.bbl}{} 
%
% Without BibTeX 

\end{article} 
\end{document}